\documentclass[aps,pre,amsmath,amssymb,citeautoscript,reprint]{revtex4-2}
\usepackage{graphicx, float, xcolor, pgf}
\usepackage{physics, siunitx}
\usepackage[caption=false, justification=centerlast]{subfig}

\makeatletter
\let\oldtheequation\theequation
\renewcommand\tagform@[1]{\maketag@@@{\ignorespaces#1\unskip\@@italiccorr}}
\renewcommand\theequation{(\oldtheequation)}
\makeatother

\usepackage{hyperref}
\hypersetup{
  pdfauthor={Amartya Bose},
  pdftitle={Pairwise Connected Tensor Network Path Integral},
  colorlinks=true,
  linkcolor=black,
  citecolor=black,
  urlcolor=black
}

\begin{document}
\title{A Pairwise Connected Tensor Network Representation of Path Integrals}
\author{Amartya Bose}
\affiliation{Department of Chemistry, Princeton University, Princeton, New Jersey 08544}
\allowdisplaybreaks

\begin{abstract}
  It has been recently shown how the tensorial nature of real-time path
  integrals involving the Feynman-Vernon influence functional can be utilized
  using matrix product states, taking advantage of the finite length of the
  non-Markovian memory. Tensor networks promise to provide a new, unified
  language to express the structure of path integral. Here, a generalized tensor
  network is derived and implemented specifically incorporating the pairwise
  interaction structure of the influence functional, allowing for a compact
  representation and efficient evaluation. This pairwise connected tensor
  network path integral (PCTNPI) is illustrated through applications to typical
  spin-boson problems and explorations of the differences caused by the exact
  form of the spectral density. The storage requirements and performance are
  compared with iterative quasi-adiabatic propagator path integral and iterative
  blip-summed path integral. Finally, the viability of using PCTNPI for
  simulating multistate problems is demonstrated taking advantage of the
  compressed representation.
\end{abstract}
\maketitle

\section{Introduction}
Tensor networks (TN) are designed to be compact ``factorized'' representations
of high-ranked tensors. Probably the most common use of TN in physics is related
to representations of the quantum many-body wavefunction which, in general, is
also a high-ranked tensor. This use has been widely demonstrated in a multitude
of methods such as the density matrix renormalization group
(DMRG)~\cite{White1992,Schollwock2005} which uses a Matrix Product State
(MPS)~\cite{Schollwock2011,Schollwock2011b} representation, and
multi-configuration time-dependent Hartree (MCTDH)~\cite{Beck2000} and its
multi-layer version (ML-MCTDH)~\cite{Wang2003, Schulze2016, Shibl2017a} which
use tree tensor networks. For multidimensional systems, an ``extension'' of MPS
to multiple dimensions called projected entanglement pair states
(PEPS)~\cite{Orus2014} is used. For systems at critical points, an MPS
representation does not work because of long-range correlations necessitating
the use of the so-called multi-scale entanglement renormalization ansatz
(MERA)~\cite{Vidal2007b,Vidal2008}. Tensor networks, since its introduction,
have proliferated in various diverse fields requiring the use of compact
representations of multidimensional data like machine learning and deep neural
networks.

While quantum dynamics at zero temperature can often be simulated using
wave-function based methods like time-dependent
DMRG~\cite{White2004,Schollwock2005,Ma2018} or MCTDH, at finite temperatures,
owing to the involvement of a manifold of vibrational and low frequency
ro-translational states in the dynamics, they suffer from an exponentially
growing computational requirements. Feynman's path integral provides a very
convenient alternative for simulating the time-dependent reduced density matrix
(RDM) for the system. The vibrational states of the ``solvent'' introduced as
harmonic phonon modes under linear response~\cite{Makri1999} are integrated out
leading to the Feynman-Vernon influence functional~\cite{Feynman1963}. Identical
influence functional also arises in dealing with light-matter interaction
through the integration of the photonic field.

The primary challenge in using influence functionals and path integrals is the
presence of the non-local history-dependent memory that leads to an exponential
growth of system paths. While many recent developments have helped improve the
efficiency of
simulations~\cite{Makri2012,Makri2014,Makri2014a,Makri2020,Makri2020a}, each of
them utilize very different and deep insights into the structure of path
integrals. It has recently been shown that the MPS representation can be very
effectively utilized to reformulate real-time path integrals involving the
influence functional leveraging the finite nature of the non-local
memory~\cite{Strathearn2018,Guo2020,Fux2021,Bose2021b}. While the MPS structure
is the simplest tensor network that can be used, the 1D topology is probably not
optimal when the non-Markovian memory spans a large number of time-steps and
suffers from growing bond dimensions. In this paper, an alternate generalized
tensor network that directly captures the pairwise interaction structure of the
Feynman-Vernon influence functional, is introduced. This pairwise connected
tensor network path integral (PCTNPI) has an extremely compact representation,
that can be efficiently evaluated, allowing us to go to much longer
non-Markovian memories without resorting to various techniques of path
filtration. Tensor networks show great promise in being a unifying language for
formulating and thinking about path integral methods.

The construction and evaluation of the tensor network is discussed in
Sec.~\ref{sec:method}. In Sec.~\ref{sec:results}, we illustrate some typical
applications of the algorithm. The memory usage is also reported for various
parameters. The implementation of this method utilized the open-source
ITensor~\cite{ITensor} library for tensor contractions allowing for extremely
efficient tensor contractions using highly efficient BLAS and LAPACK libraries.
We end the paper in Sec.~\ref{sec:conclusion} with some concluding remarks and
outlook on future explorations.

\section{Methodology}
\label{sec:method}
Consider a quantum system coupled to a dissipative environment described by a
Caldeira-Leggett model~\cite{Caldeira1983,Caldeira1983a,Leggett1987}
\begin{align}
  \hat{H}                               & = \hat{H}_0 + \hat{H}_{\text{env}}\left(p, x\right)                                                         \\
  \hat{H}_{\text{env}}\left(p, x\right) & = \sum_j \frac{p^2_j}{2m_j} + \frac{1}{2}m_j \omega_j^2 \left(x_j - \frac{c_j\hat{s}}{m\omega^2_j}\right)^2
\end{align}
where $\hat{H}_{0}$ is the Hamiltonian of the $D$-dimensional system of interest
shifted along the adiabatic path~\cite{Makri1992}. If the quantum system can be
described by a two-level Hamiltonian, then
$\hat{H}_0 = \epsilon\hat{\sigma}_z - \hbar\Omega\hat{\sigma}_x$, where $\hat{\sigma}_z$ and $\hat{\sigma}_x$ are
the Pauli matrices. $\hat{H}_{\text{env}}$ represents the Hamiltonian of the
reservoir or environment modes which are coupled to some system operator
$\hat{s}$. The strength of the $j$\textsuperscript{th} oscillator is $c_j$.
While we are using a time independent Hamiltonian for simplicity,
time-dependence from an external field in the system Hamiltonian can be captured
through the corresponding system propagator in a straightforward manner.

For a problem where the environment is in thermal equilibrium at an inverse
temperature $\beta = \tfrac{1}{k_B T}$, and its final states are traced out, the
interactions between the system and the environment is characterized by the
spectral density~\cite{Caldeira1983,Makri1999}
\begin{align}
  J(\omega) & = \frac{\pi}{2}\sum_j \frac{c^2_j}{m\omega_j} \delta(\omega - \omega_j).
\end{align}
In fact, the spectral function, $S(\omega)$ corresponding to the collective bath
operator $X = -\sum_j c_j x_j$ is related to the spectral density as
follows~\cite{Leppakangas2018}:
\begin{align}
  S(\omega) & = \frac{2\hbar J(\omega)}{1 - \exp(-\beta\hbar\omega)}.
\end{align}
For environments defined by atomic force fields or \textit{ab initio}
calculations, it is often possible to evaluate the spectral density from
classical trajectory simulations.

The dynamics of the RDM of the system after $N$ time steps, if the initial state
is a direct product of the system RDM and the bath thermal density is given as:
\begin{widetext}
  \begin{align}
    \mel{s_N^+}{\rho(N\Delta t)}{s_N^-} & = \sum_{s_0^\pm}\sum_{s_1^\pm}\ldots\sum_{s_{N-1}^\pm} \bra{s_N^+}\hat{U}\dyad{s_{N-1}^+}\hat{U}\ket{s_{N-2}^+}\ldots\nonumber                      \\
                                & \times\bra{s_1^+}\hat{U}\dyad{s_0^+}\rho(0)\dyad{s_0^-}\hat{U}^\dag\ket{s_1^-}\ldots\bra{s_{N-1}^-}\hat{U}\ket{s_N^-} F[\{s^\pm_j\}] \label{eq:pi} \\
    \text{where }F[\{s^\pm_j\}]         & = \exp\left(-\frac{1}{\hbar}\sum_k(s_k^+-s_k^-)\sum_{k'\le k}(\eta_{kk'}s_{k'}^+ - \eta^*_{kk'}s_{k'}^-)\right).\label{eq:fvif}
  \end{align}
\end{widetext}
Here, $\hat{U}$ is the short-time system propagator for $\Delta t$ and $\{s_j^+\}$
and $\{s^-_j\}$ are the forward-backward system paths. The Feynman-Vernon
influence functional~\cite{Feynman1963}, $F[{s^\pm_j}]$, is dependent on the
system path $s_j^\pm$ and the bath response function that is discretized as the
$\eta_{kk'}$-coefficients~\cite{Makri1995_1,Makri1995_2}. The influence functional
depends upon the history of the system path, leading to the well-known
non-Markovian nature of system-environment decomposed quantum dynamics. Notice
that it can be factorized based on the ``range'' of interaction in the following
manner:
\begin{align}
  F[\{s_j^\pm\}]             & = \prod_{\alpha=0}^{N} \prod_{k=\alpha}^N I^{(\alpha)}_{s_k, s_{k-\alpha}} \\
  I^{(\alpha)}_{s_{k'}, s_k} & =
  \exp\left(-\frac{1}{\hbar}(s^+_k - s^-_k)(\eta_{kk'}s^+_{k'} - \eta_{kk'}^*s^-_{k'})\right)\delta_{k', k-\alpha}.
\end{align}

The influence functional creates pairwise interactions between points that are
temporally separated. As it has been shown, if MPS and MPO are used to model the
influence functional, the fact that these interactions can spread across long
temporal spans leads to an increase in the effective bond dimension. Here, the
goal is to create a structure that naturally and efficiently accounts for the
pairwise interactions that span long temporal separations while not being
associated with any one particular representation.

\begin{figure}
  \includegraphics[scale=0.6]{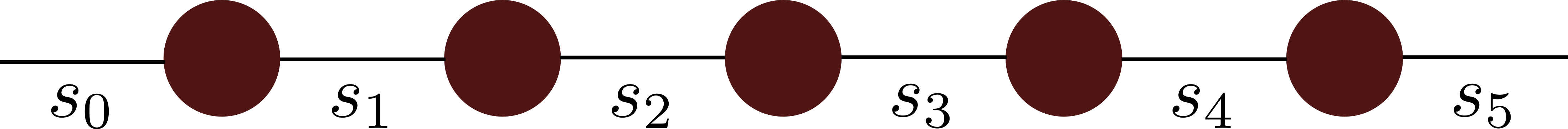}
  \caption{Diagram for $P^{(1)}$ for a $5$-step propagation. Dark brown circles
    represent the $K$ tensors.}\label{fig:mps}
\end{figure}
To motivate the tensor network representation, first consider the Markovian part
of Eq.~\ref{eq:pi}, involving just the propagators and the terms of the
influence functional coupling consecutive time points. These terms can be simply
rearranged as:
\begin{align}
  P^{(1)}_{s_0,s_N}            & = K_{s_0,s_1}K_{s_1,s_2}\ldots K_{s_{N-1},s_N}\label{eq:mark}                                 \\
  K_{s_j,s_{j-1}} & = \mel{s_j^+}{\hat{U}}{s_{j-1}^+}\mel{s_{j-1}^-}{\hat{U}^\dag}{s_j^-} I^{(1)}_{s_{j-1}, s_j} I^{(0)}_{s_{j}, s_{j}},\text{ }j\ge2\\
  K_{s_1,s_0} & = \mel{s_1^+}{\hat{U}}{s_0^+}\mel{s_0^-}{\hat{U}^\dag}{s_1^-} I^{(1)}_{s_0, s_1} I^{(0)}_{s_{1}, s_{1}} I^{(0)}_{s_{0}, s_{0}}.
\end{align}
Here, we are implicitly summing over repeated indices that do not appear on both
sides of the equation. The $\pm$ labels on the site indices of the tensors are
omitted for convenience of notation. The superscript, 1, on $P$ is there to
denote the maximum distance of interaction that we have incorporated.
Equation~\ref{eq:mark} is already a tensor network; more specifically it is
series of matrix multiplication as shown in Fig.~\ref{fig:mps}. Let us now bring
the ``next-nearest neighbor'' interactions $I^{(2)}_{s_{j-2}, s_j}$. Clearly, it
is not possible to directly contract the $I^{(2)}_{s_{j-2}, s_j}$ tensor to the
$P^{(1)}$ tensor because the internal $s_j$'s have already been traced over. To
make it possible to incorporate the $I^{(2)}$ tensors, we augment the $K$
tensors as follows:
\begin{align}
  \mathbb{K}^{r_1}_{s_0,s_1}               & = K_{s_0,s_1}\delta_{s_0,r_1}                                                                     \label{eq:aug1} \\
  \mathbb{K}^{l_{N-1}}_{s_{N-1},s_N}       & = K_{s_{N-1}, s_N}\delta_{s_N, l_{N-1}}                                                           \label{eq:aug2} \\
  \mathbb{K}^{l_{j-1}, r_j}_{s_{j-1}, s_j} & = K_{s_{j-1}, s_j} \delta_{s_j,l_{j-1}} \delta_{s_{j-1}, r_j}\quad\text{if }j\ne 1\text{ and } N.\label{eq:aug3}
\end{align}

\begin{figure}[h]
  \includegraphics[scale=0.6]{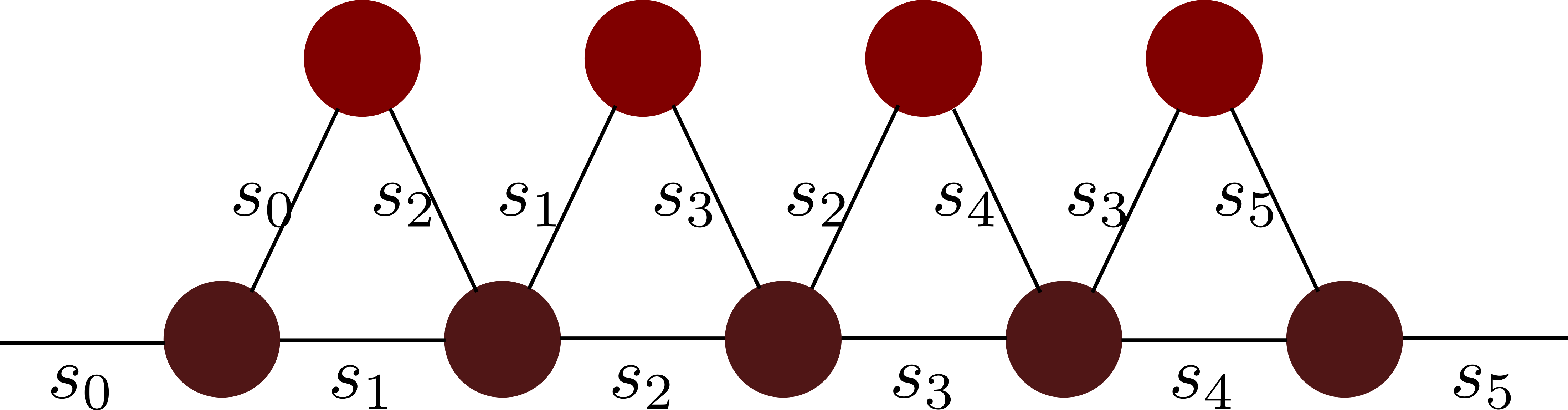}
  \caption{Diagram for $P^{(2)}$ for a $5$-step propagation. The darker circles
    which form the base represent the $\mathbb{K}$ tensors. The comparatively
    lighter red circles forming the second layer represent the $I^{(2)}$
    tensors. The labels of the non-horizontal indices have been reverted back to
    the $s$ coordinate by using the $\delta$ function relations in
    Eq.~\ref{eq:aug1}, Eq.~\ref{eq:aug2} and Eq.~\ref{eq:aug3}.}\label{fig:P_2}
\end{figure}

It is convenient to think of lower indices as the ``input'' indices and the
upper indices as the ``output'' indices, though there is no other mathematical
significance to the positioning of the indices. With this input-output
convention in mind, it is easy to see that the internal augmented $\mathbb{K}$
tensors duplicate and flip the order of the input indices, $s_j$. This ensures
that indices that differ by two time steps are now placed adjacent in the output
layer. Now, the Markovian terms and the $I^{(2)}$ terms can be combined and we
get:
\begin{align}
  P^{(2)}_{s_0,s_N} & = \mathbb{K}^{r_1}_{s_0,s_1}\prod_{j=1}^{N-2} I^{(2)}_{r_j,l_j} \mathbb{K}^{l_j,r_{j+1}}_{s_j,s_{j+1}} I^{(2)}_{r_{N-1},l_{N-1}}\mathbb{K}^{l_{N-1}}_{s_{N-1}, s_N},
\end{align}
which is depicted in Fig.~\ref{fig:P_2}. Notice that the index $s_j$ still
connects $\mathbb{\tilde{K}}_{s_{j-1}, s_j}$ and $\mathbb{\tilde{K}}_{s_j,
  s_{j+1}}$, as in $P^{(1)}$, but now there is another connection that goes
through the $I^{(2)}$ tensor in a ``triangular'' form. This feature of an
$I^{(\alpha)}$ with a higher $\alpha$ acting as a bridge between
$\mathbb{\tilde{K}}$ or $I$ tensors with smaller values of $\alpha$ would become
a recurring motif in this tensor network.

\begin{figure}
  \includegraphics[scale=0.6]{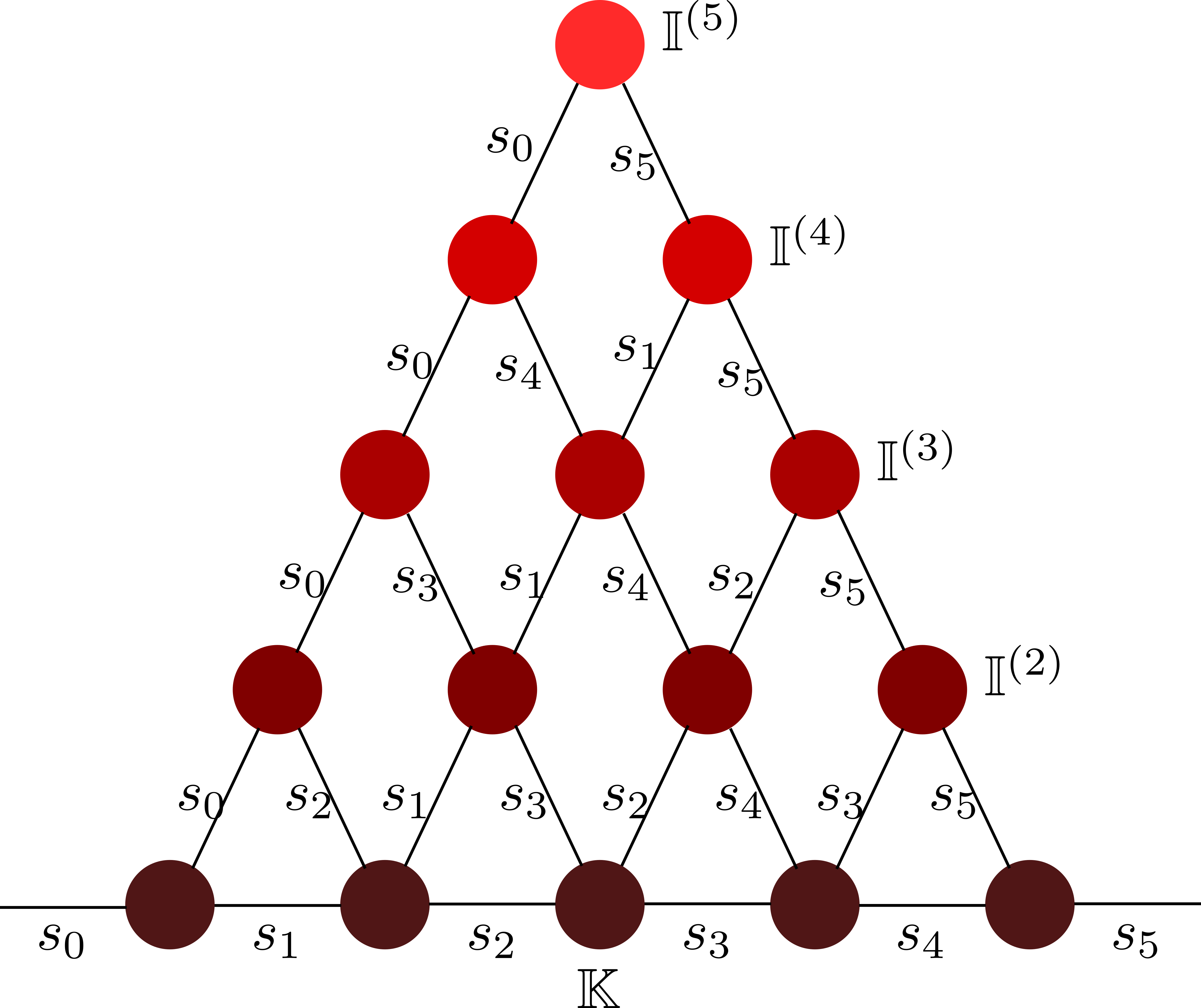}
  \caption{Diagram for the final Green's function for a $5$-step propagation.
    Dark brown circles represent the $\mathbb{K}$ tensors. The various red
    circles represent the $\mathbb{I}$ tensors for different separations. The
    lighter reds show a larger separation between interacting
    time-points.}\label{fig:P_f}
\end{figure}

The pattern for inclusion of the rest of the non-local interactions is quite
similar. Note that in Fig.~\ref{fig:P_2}, if we did the same ``trick'' of
duplicating and flipping the order of the inputs, in the next layer indices that
differ by three time points, like $s_{0}$ and $s_{3}$, $s_{1}$ and $s_{4}$, are
going to be adjacent. Hence, this can now be multiplied by
$I^{(3)}_{s_{j-3}, s_j}$. Continuing like this, we can complete the network. The
diagram is shown in Fig.~\ref{fig:P_f}~\cite{StrathearnThesis2020}, these
augmented tensors are going to be written as $\mathbb{K}$, $\mathbb{I}^{(2)}$,
\ldots. The tensor network shown in Fig.~\ref{fig:P_f}, which we will conventionally
denote by $P^{(\infty)}$, represents the final Green's function for the propagation
of the system RDM having incorporated the non-local influence from the
environment. So, $\rho(t) = P^{(\infty)}\rho(0)$.

If the system is defined to have $D$ states, then in Fig.~\ref{fig:P_f}, all the
indices have $D^2$ dimensionality corresponding to each of the possible
combination of forward-backward states. However, this is not optimal. Notice
that the influence functional tensors, $I^{(\alpha)}_{s_{k-\alpha}, s_k}$ for a
time difference of $\alpha$, depends only on the ``difference'' coordinate,
$\Delta s_k = s_k^+ - s_k^-$ of the latter time point. So, currently, we are
carrying over more information than we need to.

\begin{figure}
  \includegraphics[scale=0.6]{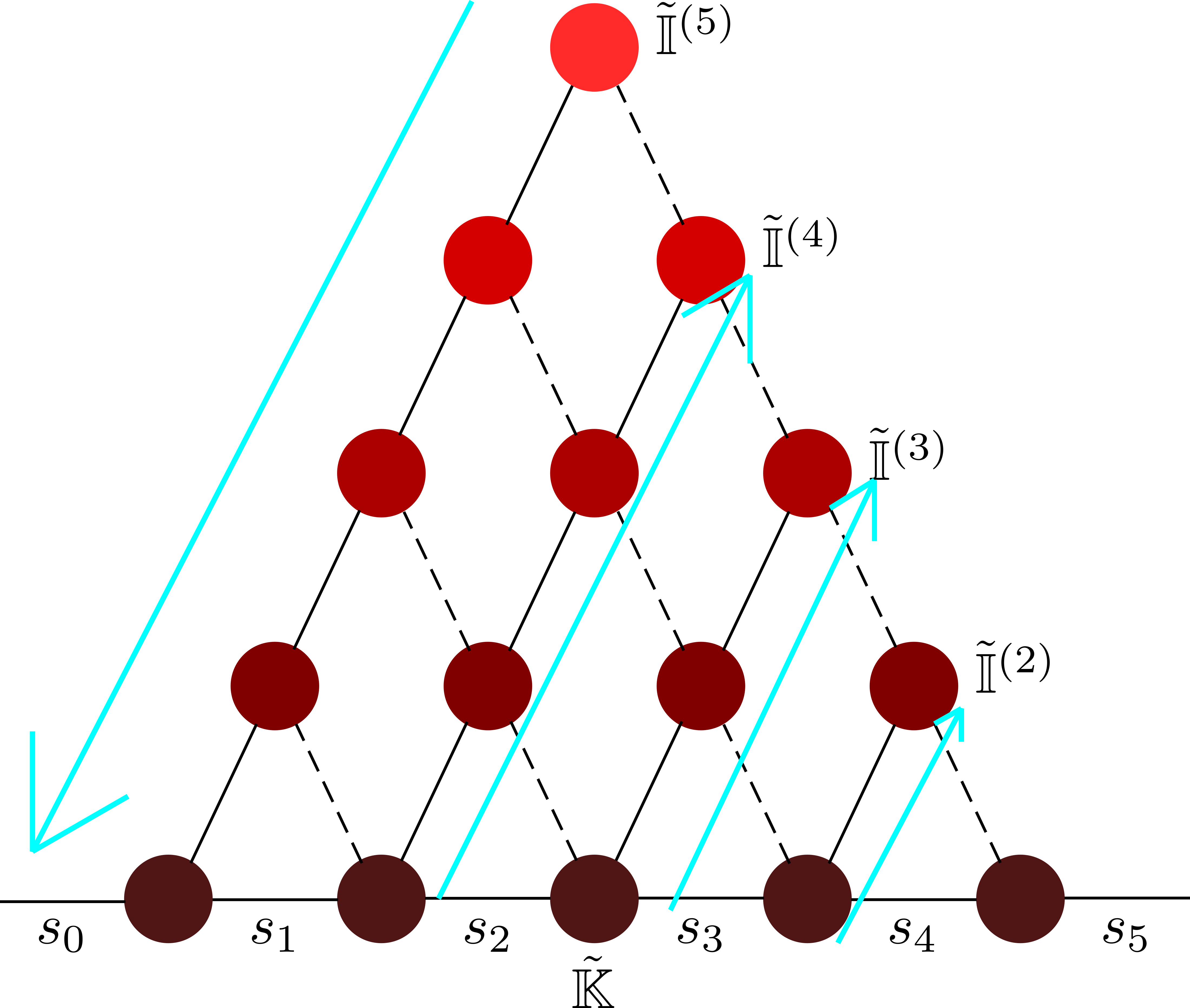}
  \caption{Optimized diagram for the final Green's function for a $5$-step
    propagation. Dashed lines have dimension $B$, and solid lines carry
    dimension $D^2$. Cyan arrows show the order of
    contraction.}\label{fig:final_contraction}
\end{figure}

To take care of this redundancy, we need to redefine the $\mathbb{K}$ tensors to
not just duplicate the input indices, but to project the ``latter'' index onto
its difference coordinates as follows:
\begin{align}
  \mathbb{\tilde{K}}^{l_{N-1}}_{s_{N-1},s_N}       & = K_{s_{N-1}, s_N}\delta_{s^+_N-s^-_N, l_{N-1}}                                                           \\
  \mathbb{\tilde{K}}^{l_{j-1}, r_j}_{s_{j-1}, s_j} & = K_{s_{j-1}, s_j} \delta_{s^+_j-s^-_j,l_{j-1}} \delta_{s_{j-1}, r_j}\quad\text{if }j\ne 1\text{ and } N.
\end{align}
Notice that the upper-right output indices in the diagrams remain exactly the
same. Only the upper-left output index of $\mathbb{K}$ changes. Therefore, the
tensor $\mathbb{K}^{r_1}_{s_0,s_1}$ remains unchanged. The dimensionality of the
``$l$'' indices is the number of unique values of $\Delta s = s^+ - s^-$ that the
system can have. For a general $D$-level system, this value is $B = D^2 - D + 1$
instead of $D^2$, however the actual symmetries present in the system might
reduce this even further. Finally, the influence functional tensors have to be
changed to be consistent, viz.
$\mathbb{\tilde{I}}^{(\alpha)}_{s_{k-\alpha}, \Delta s_k} = \exp\left(-\frac{1}{\hbar}\Delta s_k (\eta_{k, (k-\alpha)}s^+_{k-\alpha} - \eta^*_{k, (k-\alpha)}s^-_{k-\alpha})\right)$.
Even with these changes, the basic topology of the network remains the same. The
new network with the different dimensions is shown in
Fig.~\ref{fig:final_contraction}.

Having discussed the tensor network, now let us turn to the job of contracting
it. Typically, many tensor networks are constructed using singular value
decomposition (SVD) and evaluated via the truncation of the singular
values~\cite{Strathearn2018,Bose2021b}. The PCTNPI network is constructed
without resorting to any SVD calculations and consequently ``exact.'' The goal
now is to find an optimal contraction scheme that preserves this ``exactness.''
The storage cost, $S$, is also evaluated at the end of every step. The canonical
contraction order that we discuss below has been marked out in cyan arrows in
Fig.~\ref{fig:final_contraction}. For a simulation with $N$ time steps:
\begin{enumerate}
  \item Start with $\tilde{\mathbb{I}}^{(N)}_{s_0, s_N}$ and contract it with
        $\tilde{\mathbb{I}}^{(N-1)}_{s_0, s_{N-1}}$. $S = D^2 B^2$.\\
        \includegraphics[scale=0.4]{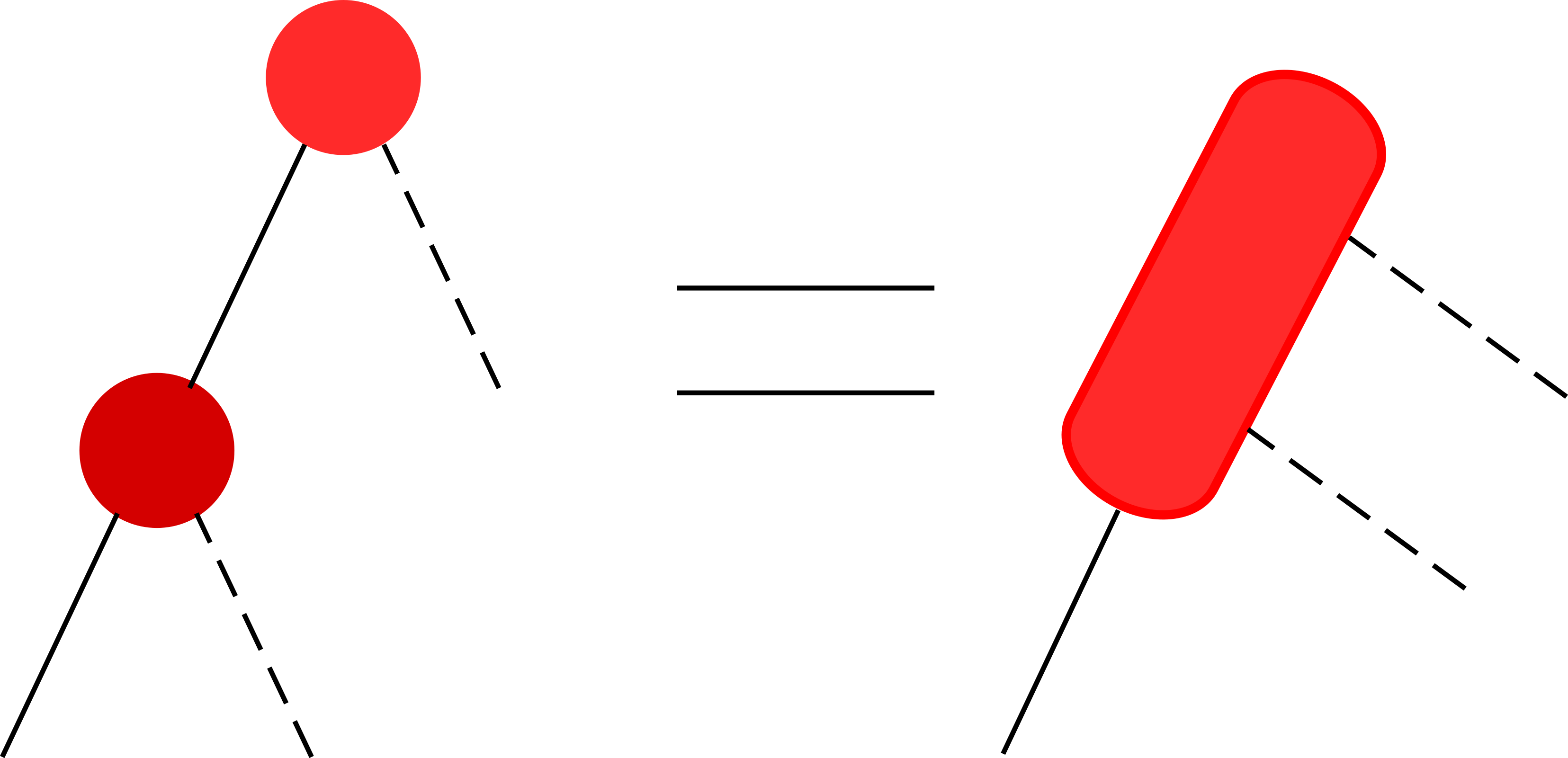}
  \item Multiply by $\tilde{\mathbb{I}}^{(N-2)}_{s_0,s_{N-2}}$. $S = D^2 B^3$.
  \item Multiply all $\tilde{\mathbb{I}}^{(\alpha)}_{s_0, s_\alpha}$ followed by
        $\mathbb{\tilde{K}}_{s_0,s_1} \rho_{s_0}$. At this stage the storage
        cost is $S = D^2 B^{N-1}$.
  \item Contract the second edge sequentially, starting from
        $\mathbb{\tilde{K}}_{s_1,s_2}$. $S = (D^2)^2 B^{N-2}$.
  \item While contracting the remaining $N-3$ tensors on the second edge, the
        storage cost remains constant at $S = (D^2)^2 B^{N-2}$.
  \item Lastly, the topmost tensor on the second edge needs to be contracted.
        The storage drops to $S = D^2 B^{N-2}$.
  \item Continuing in the same fashion, the storage requirements of contracting
        the internal tensors of the $j$\textsuperscript{th} edge is $S = (D^2)^2
          B^{N-j}$ when $j<N$.
  \item After contracting the final tensor on the $j$\textsuperscript{th} edge,
        the storage drops to $S = D^2 B^{N-j}$.
  \item Finally, the last tensor, $\mathbb{\tilde{K}}_{s_{N-1}, s_N}$ is
        contracted.
\end{enumerate}

In the above contraction scheme, we multiply the initial condition, $\rho_{s_0}$,
and get the final RDM. While this leads to a more efficient algorithm in terms
of the storage and computational cost, it is possible to reformulate the scheme
in terms of the Green's function by not involving the initial condition in the
contractions and evaluating $P^{(\infty)}$. An in-depth analysis of the memory and
computational cost is given in Appendix~\ref{sec:app}. Of course, the storage
requirement grows to a maximum of $(D^2)^2 B^{N-2}$ before decreasing
continuously. This na\"ive contraction scheme does not solve the problem of
storage. Still, as would be illustrated in Sec.~\ref{sec:results}, PCTNPI
outperforms both traditional iterative quasi-adiabatic propagator path integrals
(QuAPI)~\cite{Makri1995_1,Makri1995_2}, and iterative blip summed path integral
(BSPI)~\cite{Makri2014,Makri2014a}, when used without path filtration, in the
memory lengths that can be accessed without any sort of filtration. In a future
work, filtration schemes on top of PCTNPI would be introduced that can not only
deal with this problem, but would also avoid the construction and storage of the
full tensor. The focus of this paper is however on the tensor network and its
performance in the most na\"ive implementation.

It is well-known that the non-local memory of the influence functional dies away
with the distance between the points, allowing for a truncation of memory. This
idea is commonly used both in Nakajima-Zwanzig generalized quantum master
equations~\cite{Nakajima1958, Zwanzig1960, Shi2003c} and iterative
QuAPI~\cite{Makri1995_1,Makri1995_2}. In the framework of PCTNPI, the length of
the non-Markovian memory is equal to the depth of the resultant network. The
topmost tensor encodes the interaction between the most distant points, while
the bottom most tensor captures the Markovian interactions coming through the
propagator and the $I^{(1)}$ terms.

\begin{figure}
  \includegraphics[scale=0.6]{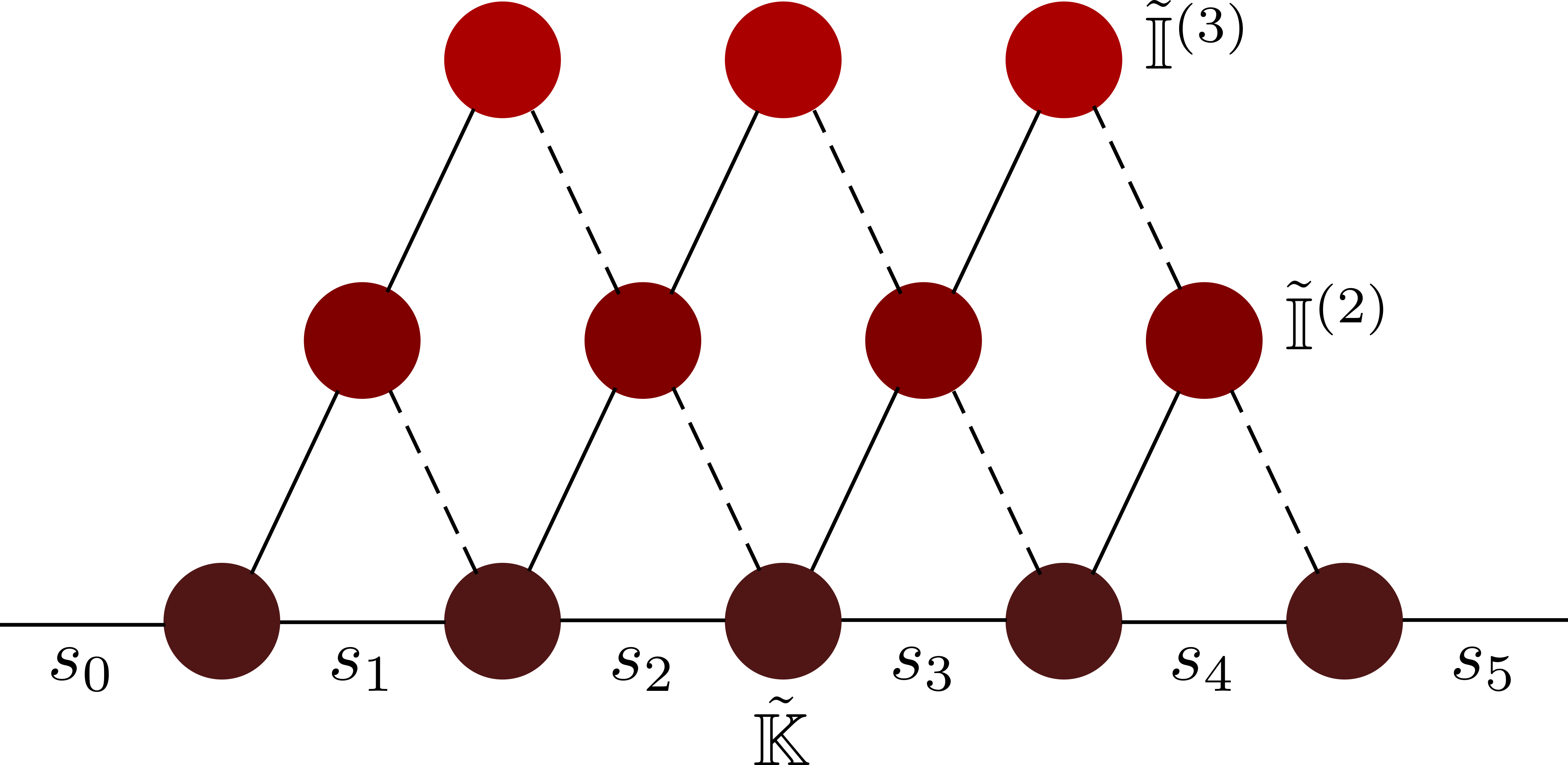}
  \caption{Diagram for the final Green function for a 5-step propagation with memory length $L=3$.}\label{fig:mem_3}
\end{figure}
\begin{figure}
  \subfloat[Contract all influence functionals with
    $s_0$.]{\makebox[0.45\textwidth][c]{\includegraphics[scale=0.6]{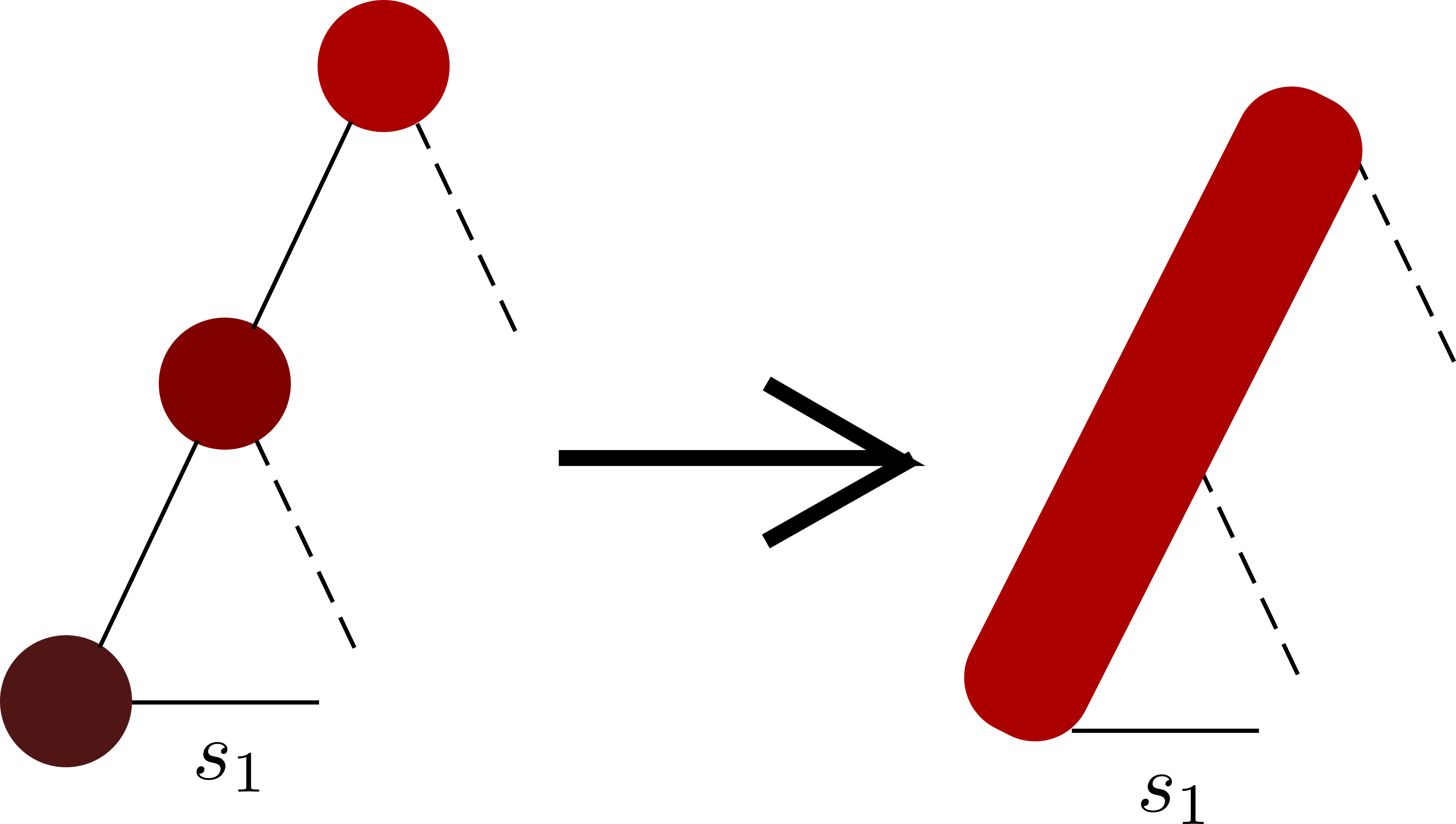}}}

  \subfloat[Contract result with all influence functionals with $s_1$. Notice
    that the external index with $s_1$ from the previous step has been
    contracted out and now the bottom-most external index is
    $s_2$.]{\makebox[0.45\textwidth][c]{\includegraphics[scale=0.6]{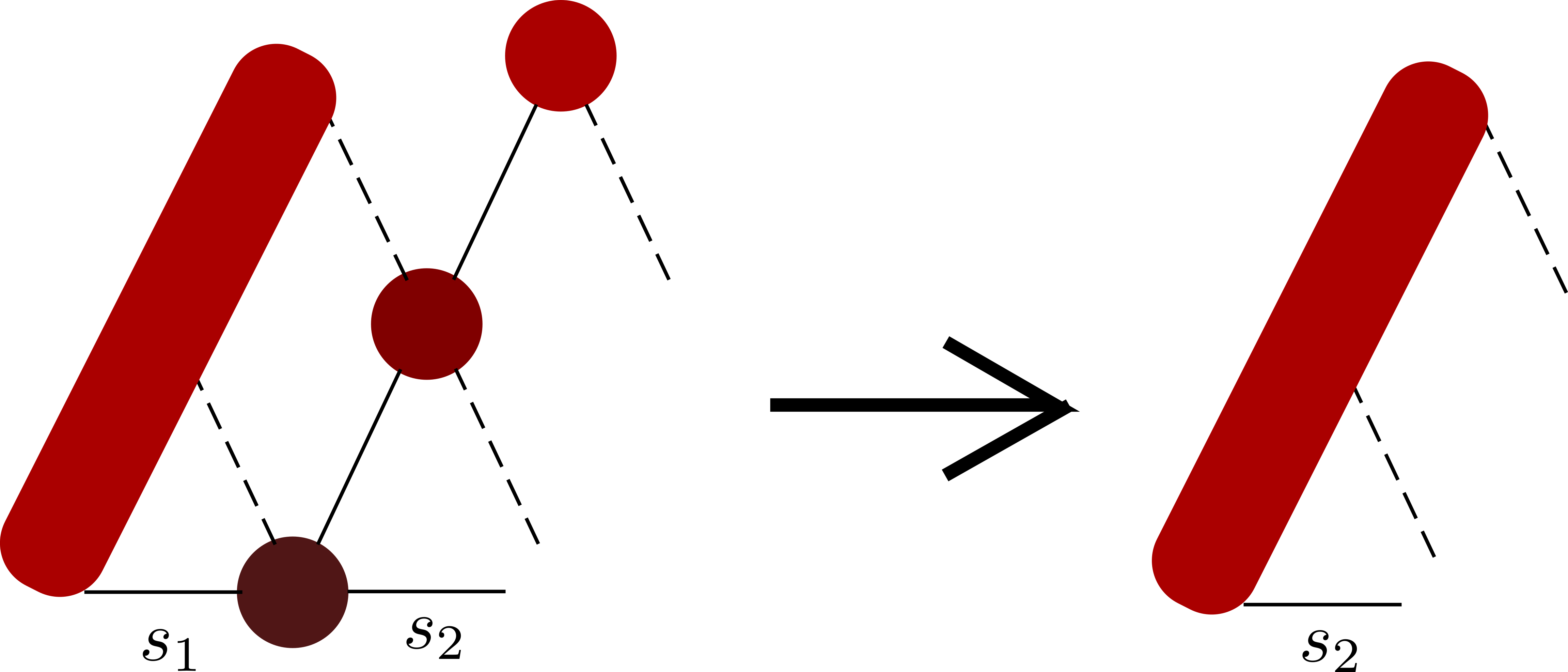}}}

  \subfloat[Contract result with remaining terms to get
    RDM.]{\makebox[0.45\textwidth][c]{\includegraphics[scale=0.6]{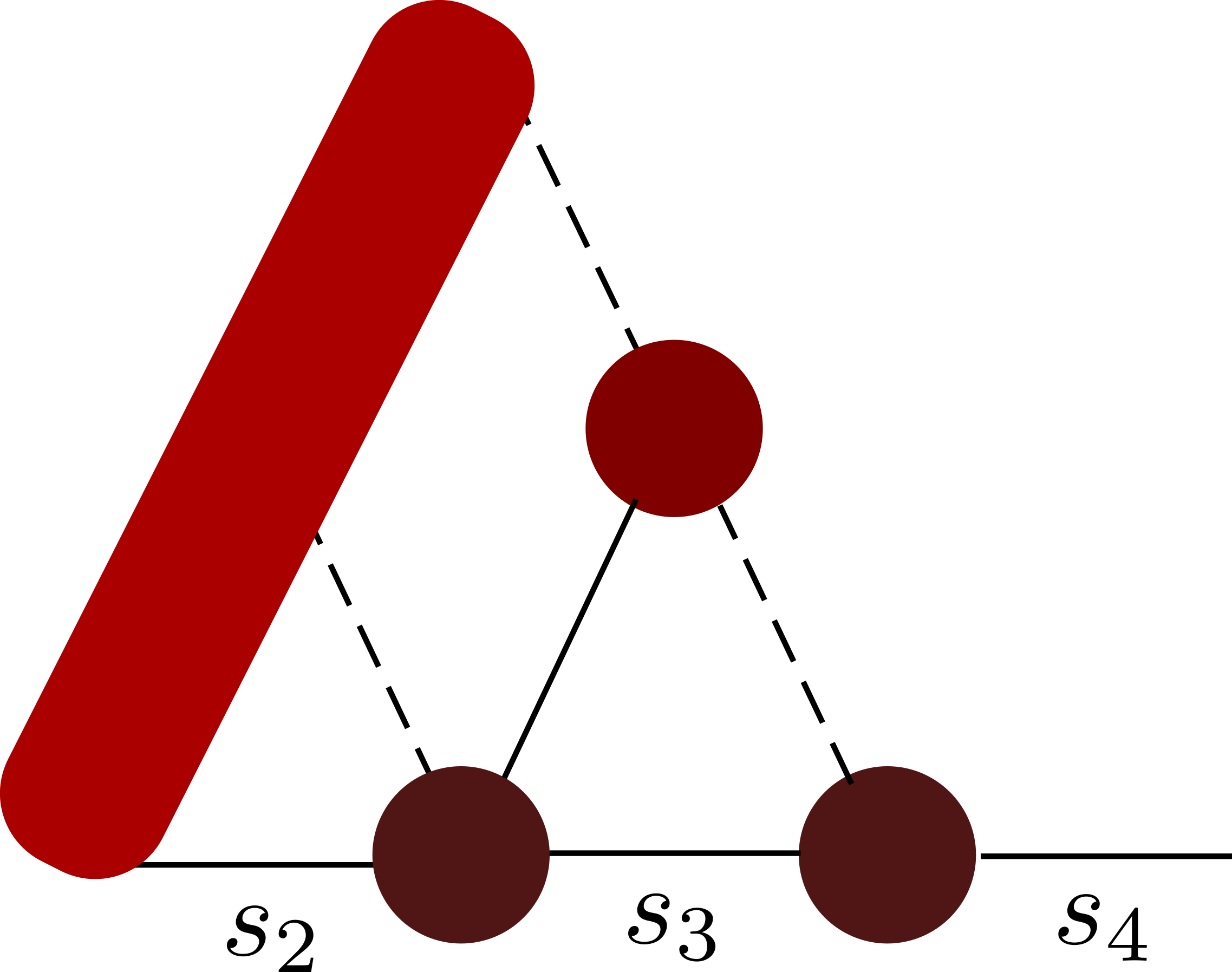}}}
  \caption{First steps of algorithm for iteration. The basic contractions are done in the same way as described for the full path part.}
  \label{fig:iter}
\end{figure}

At two time-steps of memory, that is $L=2$, we basically get Fig.~\ref{fig:P_2}.
In Fig.~\ref{fig:mem_3}, we show the structure of the network for a 5-step
propagation with $L=3$. Because $s_0$ does not interact with $s_4$ or $s_5$, it
is not necessary to store and evaluate the full diagram at once, but it can be
built iteratively. The first edge, corresponding to interactions with $s_0$ is
contracted, and multiplied by the second edge, using the canonical contraction
scheme discussed previously. As soon as this is done, the storage of the first
edge can be freed, and the third edge can be contracted. This iteration scheme
turns out to be identical to the iteration scheme in iterative QuAPI. The first
steps of the iteration algorithm is pictorally outlined in Fig.~\ref{fig:iter}.

\citeauthor{Makri2014}~\cite{Makri2014} has shown that it is possible to think
of the memory as arising from two different causes. The influence functional $F$
can be rewritten in terms of the real and imaginary parts of the
$\eta$-coefficients as:
\begin{align}
  F[\{s^\pm_j\}] & = e^{-\frac{1}{\hbar}\sum_k \Delta s_k \sum_{k'\le k}(\Re\eta_{kk'}\Delta s_{k'} - 2i\Im\eta_{kk'}\bar{s}_{k'})}
\end{align}
where $\Delta s_k = s_k^+ - s_k^-$ and $\bar{s}_k = \tfrac{1}{2}(s^+_k +
  s^-_k)$. The part of the influence functional that arises from $\Re\eta_{kk'}$
is called the classical decoherence factor. It corresponds to stimulated
phonon absorption and emission~\cite{Wang2019}. This can also be obtained
through classical trajectory-simulations and reference
propagators~\cite{Banerjee2013a} in a Markovian manner. All effects of
temperature is captured in the classical decoherence term. The term with the
$\Im\eta_{kk'}$ is the back-reaction that leads to quantum decoherence. This
part of the memory is truly non-local and temperature independent.

As a cheap approximation to the dynamics, it is possible to do a simulation with
classical decoherence, that would become increasing accurate as the temperature
of the simulation rises. In this, the full $\eta_{kk'}$ coefficients are used
only when $k=k'$ or $k=k'+1$, and otherwise the imaginary part of $\eta_{kk'}$
is ignored. (Actually, the true expressions for classical decoherence would
include the full $\eta_{kk'}$ coefficients only when $k=k'$ and the real part
otherwise. In PCTNPI, we can include the case of $k=k'+1$ as well at the same
storage and computational cost.) Effectively, we are modifying the
$I^{(\alpha)}_{s_{k'},s_k}$ operators to be $\exp\left(-1/\hbar
  \Re(\eta_{kk'})\Delta s_k \Delta s_{k'}\right)$ when $\alpha = k-k' \ge 2$. Just
like before when the $s_{k}$ lines carried unnecessary information, now the
$s_{k'}$ lines carry more information than they need to. We only need to know
about $\Delta s_{k'}$. Thus we can make the required changes to the
dimensionality of the indices by putting in the corresponding projector
operators in the $\mathbb{K}$ tensors, thereby reducing the cost of computation
even further. The network for the classical decoherence simulations would have
exactly the same structure as Fig.~\ref{fig:P_f} with all edges except the base
ones being $B$ dimensional. This approximation is especially accurate at short
times.

\section{Results}
\label{sec:results}
As illustrative examples, we apply PCTNPI to a two-level systems (TLS) coupled
bilinearly to a dissipative environment:
\begin{align}
  \hat{H}_0 & = \epsilon \hat{\sigma}_z - \hbar\Omega \hat{\sigma}_x
\end{align}
The dissipative environment is chosen to be defined by Ohmic model spectral
densities, which are especially useful in modeling the low frequency
ro-translational modes. We use the very common Ohmic form with an exponential
cutoff,
\begin{align}
  J(\omega) = \frac{\pi}{2}\hbar\xi\omega\exp\left(-\frac{\omega}{\omega_c}\right)\label{eq:spec_exp}
\end{align}
where $\xi$ is the dimensionless Kondo parameter and $\omega_c$ is the characteristic cutoff frequency and the Ohmic form with a Drude cutoff,
\begin{align}
  J(\omega) = \kappa\omega_c\frac{\omega}{\omega^2+\omega_c^2}
\end{align}
where $\kappa$ is a measure of the coupling strength. Generally these model
spectral densities are often thought to be fully characterized by a
reorganization energy
\begin{align}
  \lambda = \frac{2}{\pi}\int_{-\infty}^\infty\frac{J(\omega)}{\omega}\dd{\omega}
\end{align}
and the cutoff frequency, $\omega_c$. The reorganization energies for the exponential and the drude cutoff spectral densities are as listed below:
\begin{align}
  \lambda_{\text{Exp}}   & = 2 \xi\omega_c \\
  \lambda_{\text{Drude}} & = 2 \kappa.
\end{align}

As we demonstrate through the examples, though the reorganization energy and the
cutoff frequency are same, the exact dynamics of the reduced density matrix is
highly dependent on the form of the ``decay function.''

\begin{figure}[ht]
  \subfloat[Convergence with respect to memory length,
    $L$.]{\includegraphics{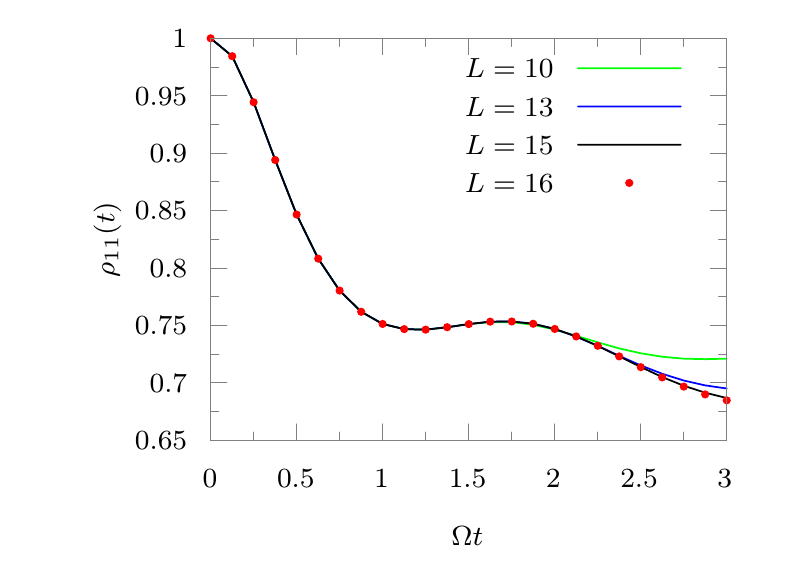}}

  \subfloat[Comparison between spectral densities with exponential and Drude
    decay functions.]{\includegraphics{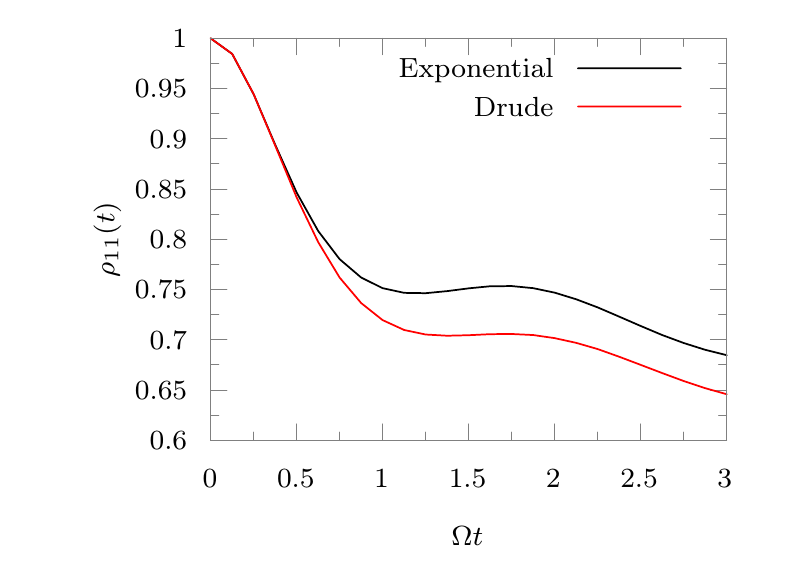}}

  \caption{Dynamics of a symmetric TLS interacting with a bath with $\lambda =
      2, \omega_c = \Omega$ at an inverse temperature $\hbar\Omega\beta =
      1$.}\label{fig:l2_w1_b1}
\end{figure}
Consider a symmetric TLS ($\epsilon = 0$) and $\Omega = 1$ interacting strongly
($\xi=2$) with a sluggish bath ($\omega_c = \Omega$) initially localized on the
populated system state $1$. The bath has a reorganization energy of $\lambda =
  4$ and is held at an inverse temperature of $\hbar\Omega\beta = 1$. The dynamics
was converged at $\Delta t = 0.125$, and a memory length $L = 16$. The
convergence is shown in Fig.~\ref{fig:l2_w1_b1}~(a) for an Ohmic bath with an
exponential decay. Full quantum-classical simulations for this parameter is
available~\cite{Walters2016}. If the Drude form of decay is used, the dynamics
changes quite significantly. The comparison between the dynamics arising from
the two spectral densities is shown in Fig.~\ref{fig:l2_w1_b1}~(b).

\begin{figure}[h]
  \subfloat[Spectral density with exponential decay
    functions]{\includegraphics{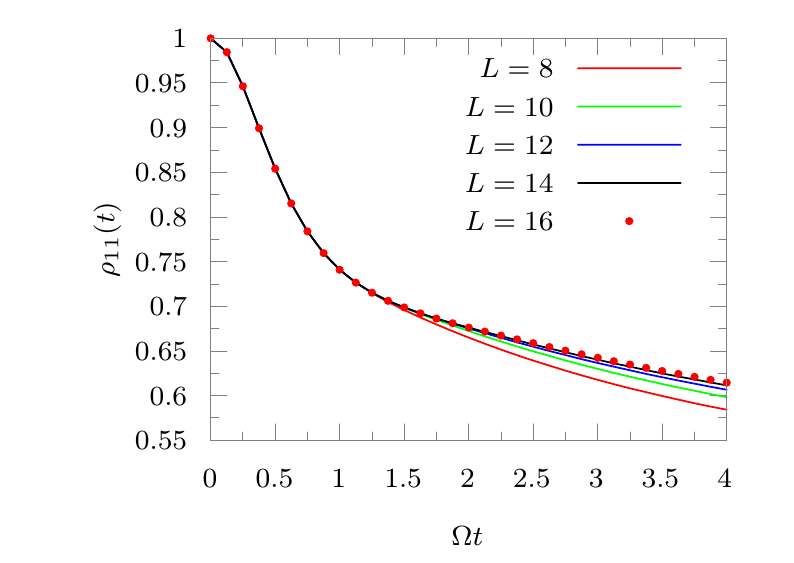}}

  \subfloat[Spectral density with Drude decay
    functions]{\includegraphics{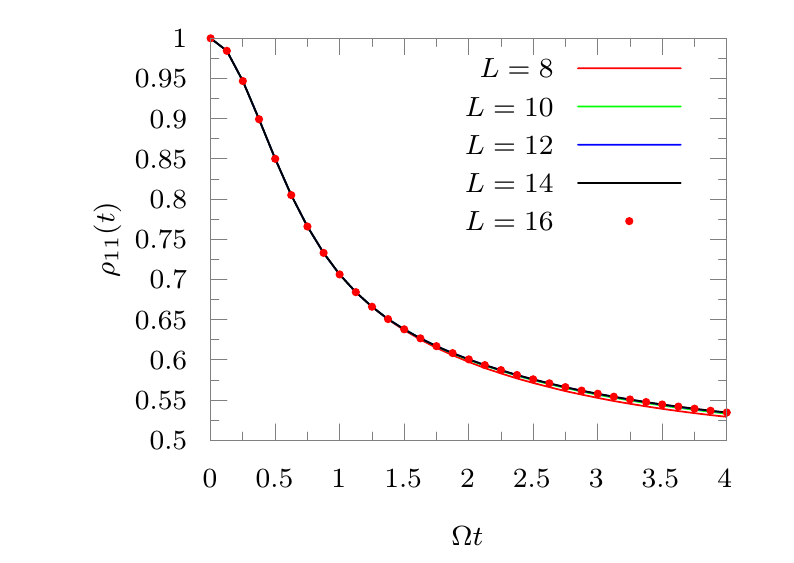}}

  \caption{Convergence with respect to memory length for the spectral density
    with an exponential cutoff.}\label{fig:l4convergence}
\end{figure}
Next, consider a case where not only is the dynamics different between the two
different decay functions, but the converged non-Markovian memory length is
different as well. The dynamics of the same TLS as above ($\epsilon = 0, \Omega
  = 1$) is now simulated in a bath with the reorganization energy $\lambda = 8$
and a characteristics cutoff frequency $\omega_c = 5$. The bath is equilibrated
at an inverse temperature of $\hbar\Omega\beta = 5$. The time-step is converged
at $\Omega\Delta t = 0.125$. The convergence of the dynamics of the reduced
density matrix on changing the memory length, $L$, is shown in
Fig.~\ref{fig:l4convergence}. While the memory length for the exponential decay
function spectral density is quite close to convergence at $L=14$, for the Drude
spectral function, it converges at $L=10$.

\begin{figure}[h]
  \includegraphics{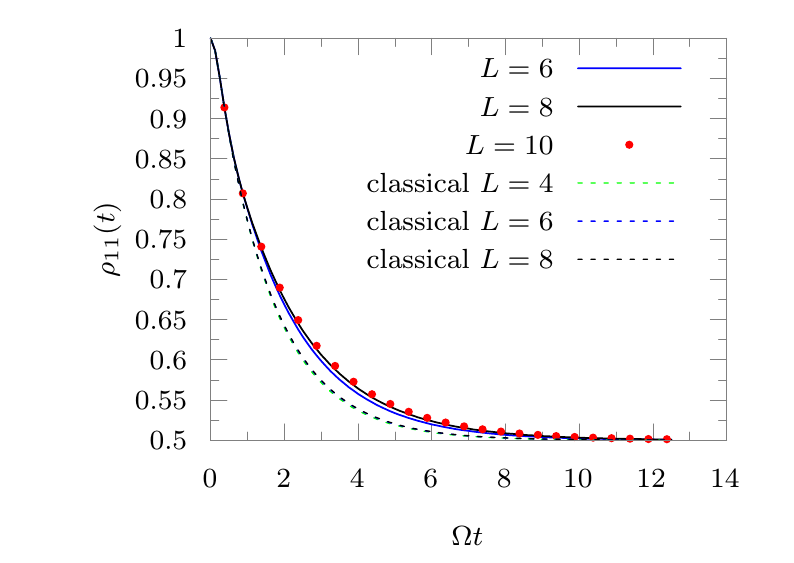}
  \caption{Comparison between the classical and full memory calculations for a
    strongly coupled high temperature bath.}\label{fig:classmem}
\end{figure}
In Fig.~\ref{fig:classmem}, we consider a TLS coupled to a strongly coupled
Ohmic bath with an exponential cutoff ($\xi = 1.2$, $\omega_c = 2.5\Omega$) equilibrated at
a high temperature $\hbar\Omega\beta = 0.2$. The converged time step is $\Delta t = 0.125$. The
classical memory calculations converge at a comparatively lower memory length,
$L$ and agree quite well with the full simulations at short times. Though at
intermediate and long times, the classical decoherence dynamics differs from the
true dynamics, this can often be enough for estimating timescales of processes,
especially using rate theory~\cite{Miller1974,Miller1983,Topaler1993a,Bose2017}.

\begin{figure}[h]
  \subfloat[Plot of memory requirements with respect to memory length for
    different methods.]{\includegraphics{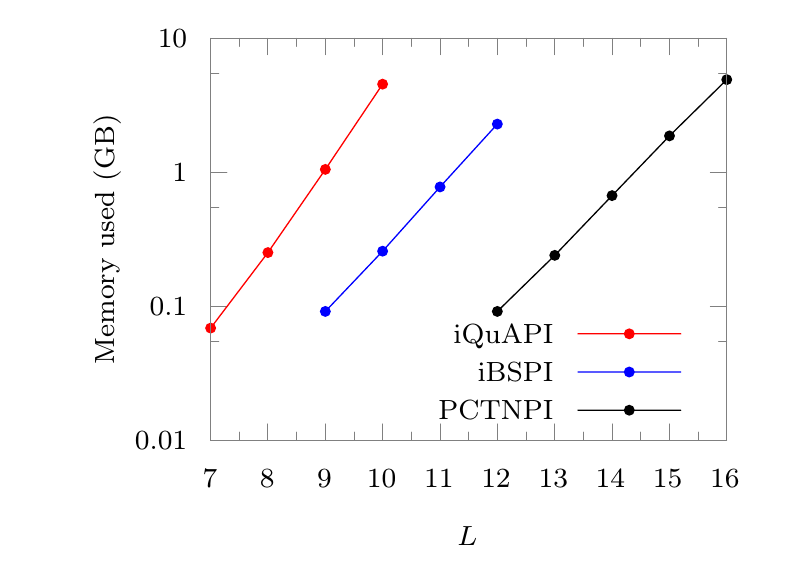}}

  \subfloat[Plot of execution times with respect to memory length for different
    methods.]{\includegraphics{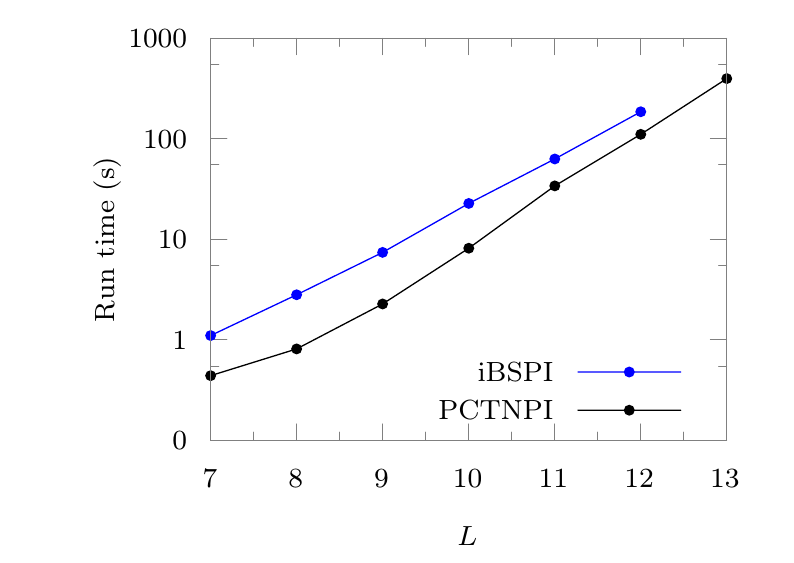}}

  \caption{Comparison of PCTNPI with other methods. Note that while the memory
    used is independent of machine, but the execution time is
    not.}\label{fig:comp}
\end{figure}
Next, the storage requirements of PCTNPI is compared with that of iterative
QuAPI and iterative BSPI in Fig.~\ref{fig:comp}~(a). To keep the comparisons
fair, the iQuAPI and iBSPI methods were run without any path filtering. It is
quite clear from the plot that the scaling of PCTNPI is essentially ``like''
that of iBSPI, i.e. for a TLS, $3^L$ scaling for iBSPI and PCTNPI vs $4^L$
scaling of iQuAPI. However, the prefactor is much smaller, allowing us to access
much longer memories with limited resources. In fact, this difference in the
prefactor would grow with the dimensionality of the quantum system. In iBSPI,
there would be $3^L$ paths for a memory length of $L$, but the storage is more
than just a number corresponding to each path. It stores a small dimensional
matrix for each path. This is the cause of the larger prefactor.

A comparison of the run times of PCTNPI with respect to iBSPI without any
filtration for a simulation of 100 time steps is presented in
Fig.~\ref{fig:comp}~(b). A laptop with Intel\textregistered{}
Core\textsuperscript{TM} i5-4200U CPU with a clock speed of 1.60GHz was used for
these benchmark calculations. These measurements are not going to be consistent
with similar benchmarks run on other machines, but the basic trends would
continue to hold. The PCTNPI algorithm is built on top of ITensor and
automatically uses parallel BLAS and LAPACK wherever possible. There is no
standard iBSPI code. The iBSPI program used for these benchmarks was manually
parallelized with OpenMP loop parallelization.

As a final example, consider a molecular wire described by the tight-binding
Hamiltonian involving $N$ sites:
\begin{align}
  \hat{H}_0 & = \sum_{1\le j\le N} \epsilon_j \dyad{\sigma_j} - \hbar V \sum_{1\le j<N} (\dyad{\sigma_j}{\sigma_{j+1}} + \dyad{\sigma_{j+1}}{\sigma_j}).
\end{align}
The site energy of the $j$\textsuperscript{th} site is $\epsilon_j$ and the
nearest neighbor couplings are $V$. The sites are separated by unit distance
such that $\ket{\sigma_j}$ are eigenstates of the position operator,
$\hat{s}\ket{\sigma_j} = (j-1)\ket{\sigma_j}$. The site energy of all but the
first site is chosen to be zero $\epsilon_j = 0$ for $j\ne 1$ and $\epsilon_1 =
  1$. The intersite coupling is chosen to be $V = 0.025$~\cite{Lambert2012b}.

\begin{figure}[h]
  \includegraphics{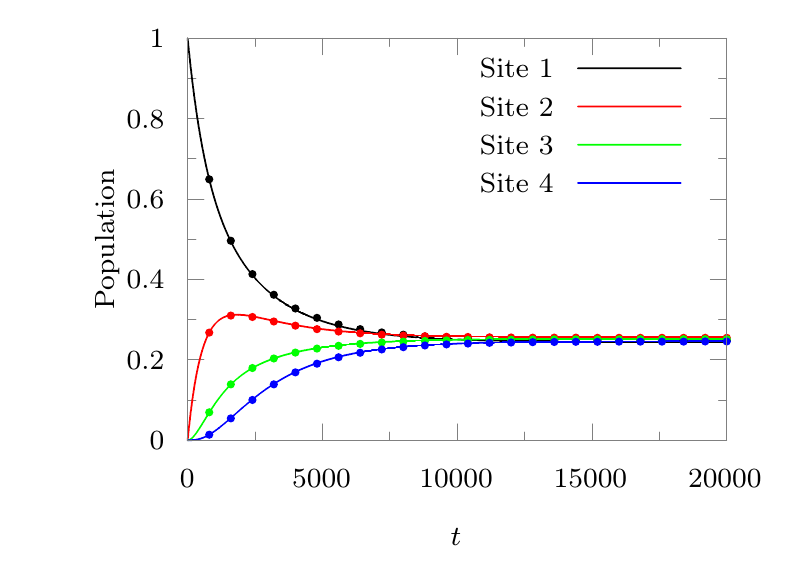}
  \caption{Population dynamics corresponding to an initially populated first
    site. Lines: full simulations, markers: classical memory simulations. For
    full decoherence, $L = 6$; for classical decoherence, $L =
      4$.}\label{fig:multidim}
\end{figure}
The computational cost grows exponentially with the number of sites. To test the
efficiency of the basic contraction scheme outlined here, we use a system with
$N = 4$ sites. The bath is characterized by an Ohmic spectral density with an
exponential cutoff, Eq.~\ref{eq:spec_exp} with $\omega_c = 4$ and $\xi =
  0.12$~\cite{Bose2021b} equilibrated at an inverse temperature of $\beta = 0.1$.
As discussed in Sec.~\ref{sec:method}, the scaling of the algorithm would go as
$B < D^2$. The symmetry of the Hamiltonian in this case ensures that the number
of unique values of $\Delta s$, $B = 7$ for this 4 state system, which is even
less than the $D^2 - D + 1$ for a completely general Hamiltonian. The population
dynamics of all the states is shown in Fig.~\ref{fig:multidim}. An initial state
with only the first site populated was used. Because of the high temperature of
the bath, the classical decoherence simulation produces practically identical
dynamics but converges at a smaller memory length $L$.

\section{Conclusion}
\label{sec:conclusion}
A novel tensor network is introduced to perform path integral calculations
involving the Feynman-Vernon influence functional. This pairwise connected
tensor network path integral (PCTNPI) captures the pairwise interaction
structure of influence functional. PCTNPI can be contracted efficiently, and
minimizes the storage requirements as far as possible without resorting to
various path filtration algorithms. Iterative decomposition of the memory is
also possible in an elegant manner. Comparisons between PCTNPI and iQuAPI and
iBSPI show the scaling of memory requirements of PCTNPI to be similar to iBSPI,
but much smaller.

PCTNPI provides an alternative to the MPS
representation~\cite{Strathearn2018,Bose2021b}, serving as a small step in
further elucidating the deep relation between tensor networks and path
integrals. While no path filtration scheme has been developed, PCTNPI is already
quite useable. It can easily incorporate classical trajectories through harmonic
backreaction quantum-classical path integrals~\cite{Wang2019} thereby making it
possible to include anharmonic effects of the environment in an approximate
manner without any additional cost. Additionally, harmonic backreaction also
leads to an increase in the converged time-step and a decrease in the effective
memory length such that some ultrafast reactions can be simulated directly.
Taking advantage of the extended memories that are accessible with PCTNPI, the
combined method would be able to simulate systems with strongly coupled sluggish
realistic solvents with high reorganization energy. This promises to be a
fruitful avenue of research in terms of applications to electron and proton
transfer reactions.

Algorithms based on MPS representations of the augmented reduced density
tensor~\cite{Strathearn2018} or of the path-dependent Green's
function~\cite{Bose2021b} can be thought of as particular optimized
re-factorizations of the PCTNPI network. We have demonstrated the viability of
evaluating PCTNPI in a brute force manner compared to other methods. This
suggests that using the PCTNPI network directly to generate other optimized
representations might also lead to novel methods.

While ideas of path filtration were not a consideration of the present paper,
schemes based on the singular value decomposition (SVD) can be incorporated with
PCTNPI, leading to a method that significantly reduces the storage, since the
full tensor would not need to be computed and stored. This development would be
discussed in a future publication.

\section*{Acknowledgments}
I thank Peter Walters for discussions and acknowledge the support of the
Computational Chemical Center: Chemistry in Solution and at Interfaces funded by
the US Department of Energy under Award No. DE-SC0019394.

\appendix
\section{Cost of Contraction}
\label{sec:app}
\begin{figure}
  \subfloat[Contraction of internal tensor along the left edge ($j\ge 2$). $C =
      (D^2)^2 B^{N-j+1}$. $S = D^2
      B^{N-j+1}$.]{\makebox[0.45\textwidth][c]{\includegraphics[scale=0.6]{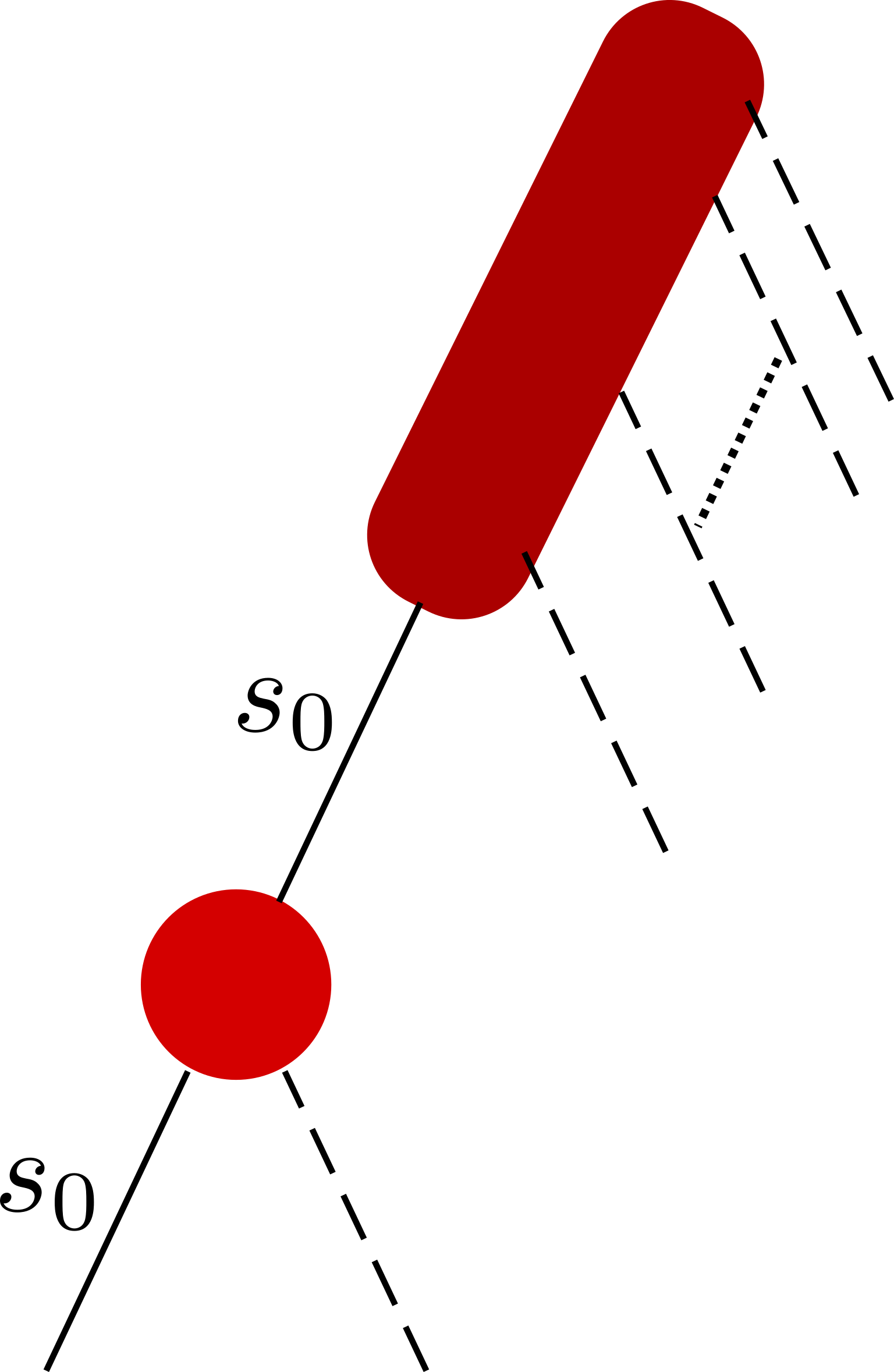}}}

  \subfloat[Contraction of $\tilde{\mathbb{K}}_{s_0,s_1}\rho_{s_0}$. $C =
      (D^2)^2 B^{N-1}.$ $S = D^2
      B^{N-1}$.]{\makebox[0.45\textwidth][c]{\includegraphics[scale=0.6]{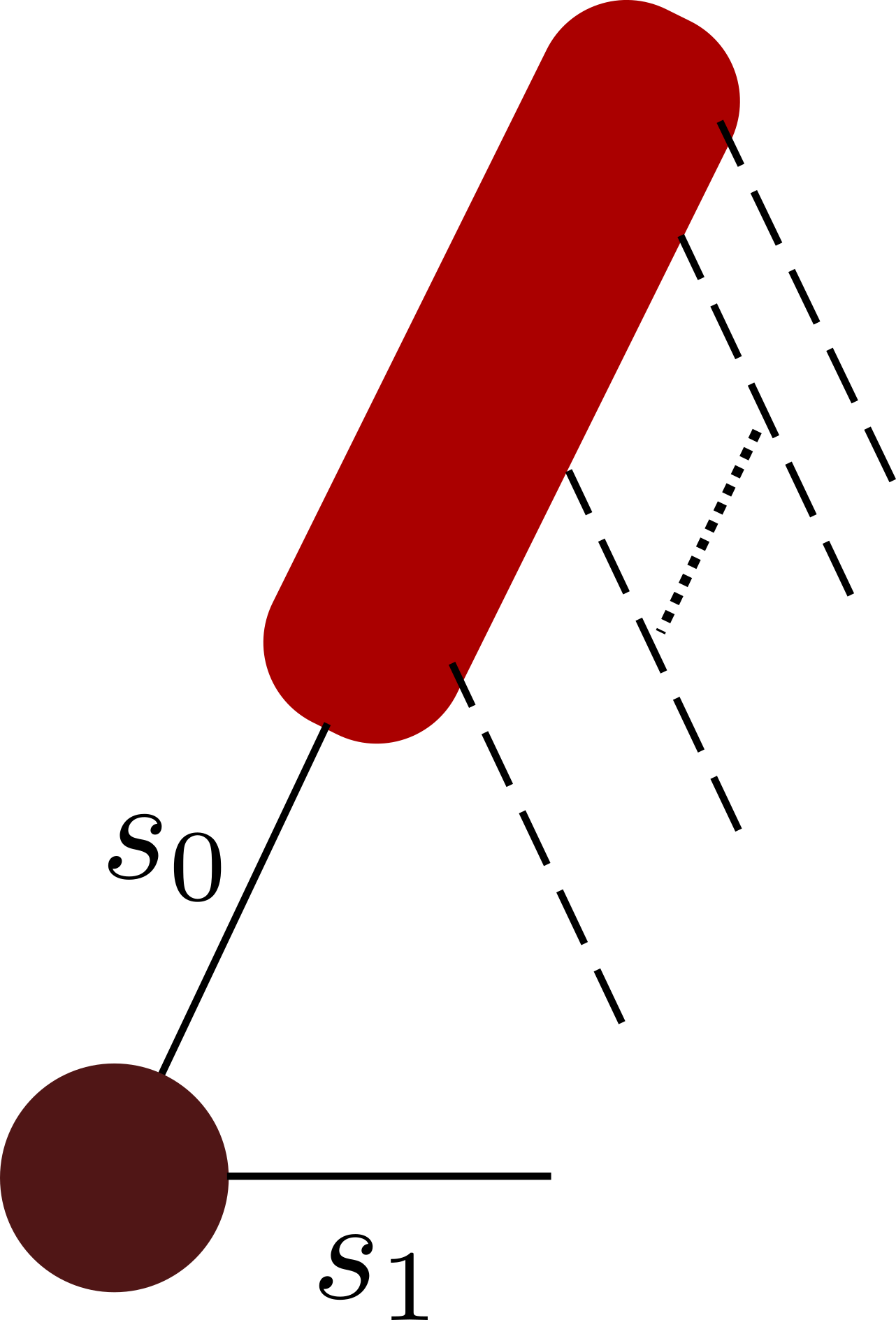}}}

  \caption{Contraction along the left edge of the triangle. Dashed lines show the $B$ dimensional indices, and solid lines show the $D^2$ dimensional indices.}\label{fig:contract_edge_gen}
\end{figure}
Consider the tensor network corresponding to a full path simulation spanning $N$
time-steps. To calculate the cost of contraction, the left ``edge'' of the
triangular network is first considered. Consider contracting
$\tilde{\mathbb{I}}^{(j)}_{s_0,s_j}$, for $j\ge 2$, with two $D^2$ indices and
one $B$ index, as schematically indicated in
Fig.~\ref{fig:contract_edge_gen}~(a). The part that has already been contracted
has one $D^2$ index and $(N-j)$ $B$ indices. Therefore, the cost of contraction
is $(D^2)^2 B^{N-j+1}$. The space requirement at this stage is $D^2 B^{N-j+1}$.
To finish the contraction of the left-most edge of the triangle, we need to
multiply by $\tilde{\mathbb{K}}_{s_0,s_1} \rho_{s_0}$ leading to the tensor
network shown in Fig.~\ref{fig:contract_edge_gen}~(b). The resultant tensor does
not have a index corresponding to $s_0$ because that has been traced over. The
computational cost of this step is $C = (D^2)^2 B^{N-1}$ and the storage becomes
$S = D^2 B^{N-1}$.

\begin{figure}
  \subfloat[Contracting the first tensor of the next parallel edge. $C = (D^2)^3
      B^{N-1}$. $S = (D^2)^2
      B^{N-2}$.]{\makebox[0.45\textwidth][c]{\includegraphics[scale=0.6]{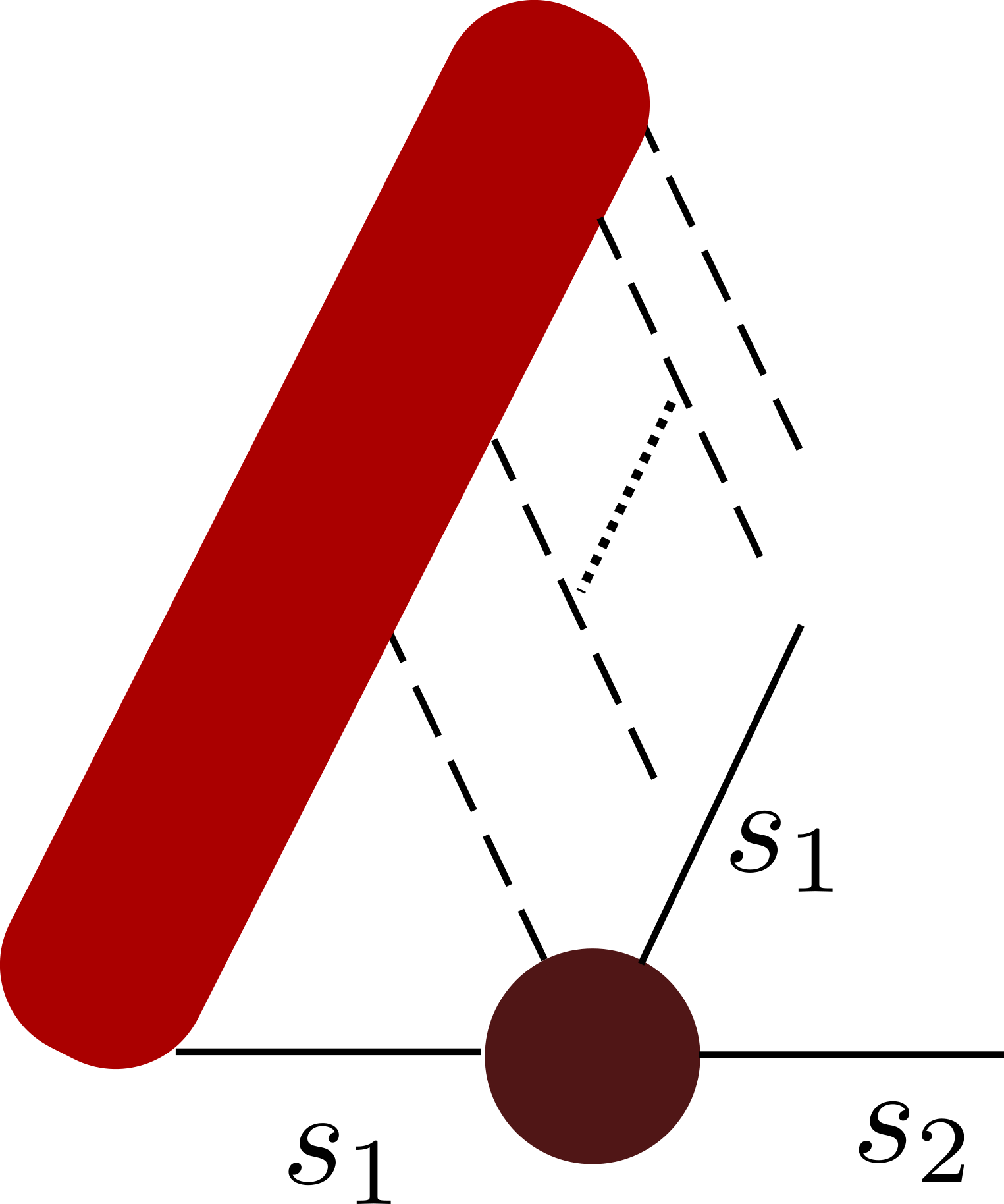}}}

  \subfloat[Contracting the last tensor of the next parallel edge. $C = (D^2)^2
      B^{N-1}$. $S = D^2 B^{N-2}$.]{\makebox[0.45\textwidth][c]{\includegraphics[scale=0.6]{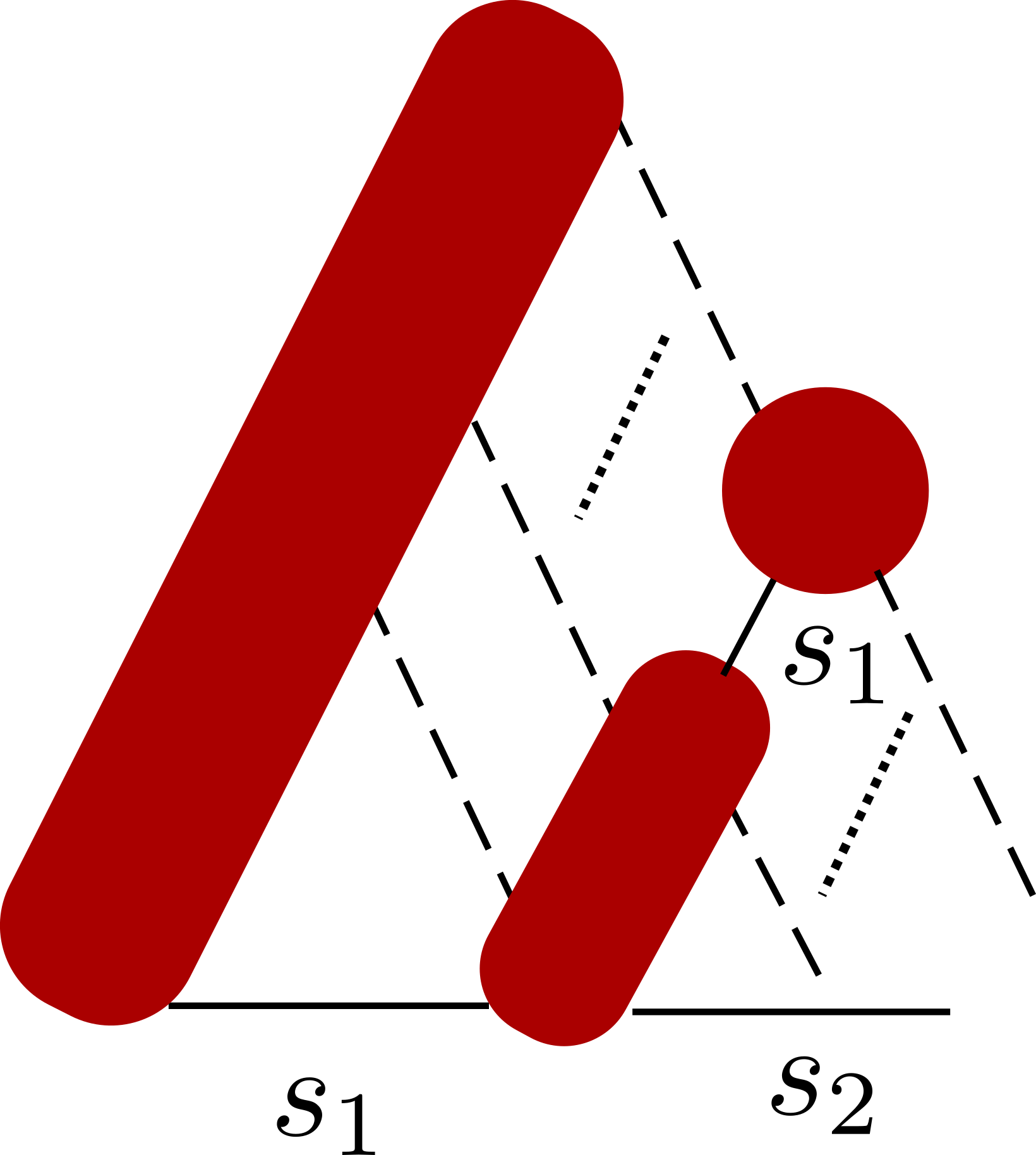}}}

  \caption{Contraction along an intermediate edge, say the one next to the
    left-most edge.}\label{fig:second-edge}
\end{figure}

Now, the second parallel edge is to be contracted. This step however is started
from the bottom, i.e. from $\tilde{\mathbb{K}}_{s_1,s_2}$. The first
contraction, shown in Fig.~\ref{fig:second-edge}~(a), is the most costly step in
the entire algorithm. The computational cost of this step is $C = (D^2)^3
  B^{N-1}$ and the storage requirement increases to $S = (D^2)^2 B^{N-2}$.
Continuing with the other intermediate tensors of the first parallel edge,
notice that the cost of contraction remains constant at $C = (D^2)^3 B^{N-1}$
and the space required remains constant at $S = (D^2)^2 B^{N-2}$. Finally, the
last, top-most tensor of this edge is to be contracted. This is illustrated in
Fig.~\ref{fig:second-edge}~(b). The computational cost is $C = (D^2)^2 B^{N-1}$.
The storage cost now drops to $S = D^2 B^{N-2}$.

Now, consider contracting a general diagonal edge, say the
$j$\textsuperscript{th} one. The resultant tensor from the previous contraction
has one $D^2$ index and $N-j+1$ $B$ indices. Contracting the
$\tilde{\mathbb{K}}$ tensor leads to a tensor with two $D^2$ indices and $(N-j)$
$B$ indices. The cost of this contraction is $C = (D^2)^3 B^{N-j+1}$ and the
storage is $S = (D^2)^2 B^{N-j}$. For all the intermediate tensors at this
stage, once again both the computational costs and the storage costs remain the
same. On contracting the last tensor of this diagonal, the storage drops to $S =
  D^2 B^{N-j}$.

Below we list the total computational cost for contracting each of the
``parallel'' edges. The edge number is given as the subscript.
\begin{align}
  C_{1} & = (D^2)^2B^{N-1} + \sum_{j=2}^{N-1} (D^2)^2 B^{N-j+1}\nonumber   \\
        & = (D^2)^2 \left(B^{N-1} + \frac{B^2 (B^{N-2} - 1)}{B - 1}\right) \\
  C_{j} & = (D^2)^2 B^{N-j+1}\left(1 + (N-j) D^2\right),\quad 2\le j\le N
\end{align}

The prefactor of the computational and storage costs is lower for classical
decoherence simulations: It goes from a power of $D^2$ to the corresponding
power of $B$. It is clear that the complexity of the entire contraction goes as
$\mathcal{O}\left(B^{N-1}\right)$ and the peak storage requirement is
$\mathcal{O}\left(B^{N-2}\right)$.

%\bibliography{/home/amartya/Documents/MendeleyDesktopLibrary/library}
\bibliography{paper.bib}
\end{document}